# Improving physico-chemical properties and antifouling of nanofiltration membranes using coating DLC nanostructures

Zeynab Kiamehr[1*], Mojtaba Shafiee[2], Babak Shokri[2, 3]

[1]Department of Physics, Tafresh University, Tafresh, Iran.
[2]Laser and Plasma Research Institute, Shahid Beheshti University, G.C, Velenjak Avenue, Tehran, Iran.
[3]Department of Physics, Shahid Beheshti University, G.C, Velenjak Avenue, Tehran, Iran.
*Email: z.kiamehr@tafreshu.ac.ir

**Abstract**

In this study for the first time, polymeric nanofiltration membranes based on polyethersulfone (PES) polymer were surface-modified by using a diamond-like carbon (DLC) nanostructure coating layer. The effect of this coating on the performance and anti-fouling properties of the membrane was investigated. The surface-modified membranes significantly improved the salt separation and their hydrophilicity was more than the control membrane after the reformation process. The amount of salt separation increased from 64% to 98% due to the decrease in the size of the holes on the surface. The absorption characteristic of DLC nanostructure in contact with feed solution was reported as another reason in this field. The contact angle (CA) of water also decreased due to the improvement of hydrophilicity, which means the increase of hydrophilicity in the above membranes. The amount of pure water flux passing through the membranes increases from one side of the membrane surface to both sides with the increase significantly of the coating level, so that its value reaches from about $8 L/m^2 h$ for the $M_0$ membrane to $65 L/m^2 h$ in the membrane containing $M_3$. The surface morphology of unmodified membranes changed from a rough state with an average of 16nm in PES membranes to a smoother state in modified membranes with an average of 9nm. The recovery rate of the flux and the total absorption rate were obtained in complete agreement with the results obtained from the CA of water and the amount of the calculated surface roughness parameter. The increase in the flux recovery rate and the decrease in the fouling rate indicated the successful improvement of the antifouling properties as a result of the coating process. Based on the washing test, the prepared sample has good stability against the number of washing cycles, which is a very important advantage for practical applications.

## Introduction

The limitation of available freshwater resources along with the increase in demand in today's industrial societies has drawn attention to wastewater treatment as well as the development of the best method to restore and reuse high-quality wastewater through advanced treatment plants. The membrane filtration process is a promising option among existing technologies [1-4]. Recently, this method has been considered an effective method to eliminate a wide range of water pollutants due to the advances in the technique and raw materials of membrane construction. According to their pore size and applied pressure, polymer membranes include several general categories of microfiltration, ultrafiltration, nanofiltration, reverse osmosis, gas separation, and evaporative seepage [5, 6]. The mechanical, chemical, and thermal stability of the membranes determines their lifespan and actually the efficiency and cost-effectiveness of the membrane filtration system. In addition, membrane clogging, which will lead to a decrease in the flux and separation performance of the membrane, causes a shorter lifespan and a higher operating cost [7, 8]. The physicochemical properties of membranes (hydrophilicity, hydrophobicity, resistance to sedimentation, chemical stability effect, etc.) have a significant impact on its performance, for example, antifouling properties are very important in the membrane separation process [9-15]. Therefore, the preparation of useful polymeric membranes with high efficiency and longevity that minimizes the need for washing and replacement can largely prevent negative environmental effects and high costs [16].

In membrane separation processes (for water purification) fouling occurs on the surface of the membrane with the passage of time, and the gradual increase in fouling leads to a decrease in the flux passing through and thus changes in the membrane's performance. Such phenomena are caused by the two factors of concentration polarization and membrane clogging [17-20]. In general, membrane clogging is a complex chemical-physical phenomenon that usually includes the deposition and absorption of clogging factors on the outer surface of the membrane as well as the inner surface of the cavities inside it [21, 22]. The clogging of the membrane holes reduces the cross-sectional area of the liquid passing through the membrane, and if the applied pressure is constant, the flux passing through the membrane decreases. In order to prevent the reduction of the throughput, the applied pressure must be increased during the separation process, which will lead to an increase in cost. Since the low operating costs in membrane separation processes are their biggest advantage, the increase in applied pressure and the subsequent increase in costs limit the performance of these membranes [23]. In general, modification of the surface characteristics of polymeric membranes is done with different methods and different goals. However, most of the surface modifications are done on polymeric membranes in wastewater treatment processes to reduce clogging [24].

Since the preparation of anti-fouling membrane was discussed, various methods have been used, including modification by grafting of hydrophilic monomers, plasma treatment, addition of nanoparticles, etc [25, 26]. Prevention of membrane fouling by providing hydrophilic polymer membranes with anti-clogging or even self-cleaning properties improves the efficiency and lifespan of the membrane and significantly reduces operating costs due to cleaning or replacing membranes. One of the most effective methods to reduce the clogging of polymer membranes is to modify the membrane using hydrophilic nanoparticles to prevent clogging and improve the separation performance of the membrane [27]. The membrane by using different methods including phase inversion methods, ([28], sol-gel, electrospinning), ([29], modification with ultraviolet rays), ([30], polymerization), ([31], Stretching and precipitation from the vapor phase can be modified. Nanoparticles, having special properties, enter the membrane structure and by changing the structure and properties of the membrane, they cause the improvement of their surface and the reduction of adhesion. Metal oxide nanoparticles such as titanium, aluminum, zirconium, silica, iron, and zinc are among the most widely used nanoparticles used in the structure of membranes [32, 33]. The main challenge in the application of titanium dioxide nanoparticles is the agglomeration of this nanoparticle in the membrane matrix and the lack of proper distribution of this nanoparticle on the surface of the membrane. By studying the previous references, the use of nanoparticles by the mixing method in the phase inversion method has created problems such as the reduction of flux due to the blocking of the holes of the polymeric membranes as a result of the non-proper distribution of these nanoparticles in the membrane matrix [34]. Li et al. studied the improvement in anti-fouling properties with the participation of nanoparticles in PES and recorded approximately 1.96% of water flux recovery (FRR) and 9.3% of irreversible sedimentation resistance [36]. Fan et al. used graphene oxide as filler in the PES membrane to improve the deposition rate [37]. Qu used the improvement in the antifouling property by adding halloysite nanotubes modified with dextran in the PES membrane mixture [38]. Deposition methods are one of the important factors in DLC coating. Currently, there are many types of deposition methods for DLC films. A wide range of temperatures (zero to 400 degrees Celsius) can be used for DLC deposition. Plating gas pressure, bias voltage, and etching time can affect the type of deposition. DLC coatings can have the desired properties if the adhesion between the coating and the substrate is strong enough [35]. DLC thin coatings have attracted the attention of many researchers due to their unique properties. These features have led to the widespread use of these coatings only in order to change the physico-chemical properties of the surface of different parts and optimize their use. DLC films mainly contain carbon in various forms such as diamond, graphite, fullerene, carbon nanotubes, and polymer. The deposition method, the amount of hydrogen, and the amount of doping determine the amount of SP2 and SP3 bonds.

Due to the practical and abundant use of nanofiltration polymer membranes, making them hydrophilic/hydrophobic and thus strengthening their anti-fouling properties has been the goal of many studies in the field of membrane processes in recent years and significant progress has been made in this field. In this article, we have tried to provide a new view of plasma modification of hydrophobic nanofiltration membranes (PES). Experience has proven that in filtering processes, hydrophobic membranes are much more prone to clogging than hydrophilic membranes. For this reason, for the first time, surface modification of the PES membrane using DLC nanostructures on both sides of the membrane, separately and simultaneously, was carried out and its impressive results were described in detail below. For these hydrophobic membranes, as it was said, hydrophilicity will lead to a decrease in sediment absorption, so here, during two completely separate test stages, the surface of these membranes was modified and became hydrophilic. Finally, the antifouling performance of these modified membranes was investigated and compared with the control membrane, which determined and introduced the best approach for modifying polymer membranes and developing membrane technologies.

## Experimental

### Materials

In this research, PES membranes have been studied due to their wide use in various industries and low cost. Polyether sulfone (PES Ultrason E6020P) with a thickness of 114μm and a pore size of 0.42μm was manufactured by BASF Company, Germany. First, all the samples' surfaces are processed with argon-hydrogen plasma (with a discharge rate of 30cc), at the power of 40W and pressure of 70mTorr for 30s. The samples are generally divided into two categories, the samples that are considered for filtration (circles with a diameter of 4.7cm) and the samples that are used for testing the characteristics of the membranes (cut into 1cm x 1cm size). All samples are first dried in distilled water for 10 minutes, then in ultrasonic ethanol for 10 minutes, and then dried in a drive box. After they dry, we go to the classification of the samples, and the samples that are considered as selection control samples are separated from the other samples that are for coating and processing.

### PECVD system (Test setup)

Here, the chemical vapor deposition (PECVD: Plasma Enhanced Chemical Vapor Deposition) method is used to obtain DLC quasi-stable structure. Then, quasi-diamond carbon layers were deposited using the direct current magnetic sputtering technique with a graphite source and radio frequency sputtering. The PECVD system has a plasma reactor in which the gas particles first react with each other and then are deposited solidly on the sample's surface. $O_2$, $H_2$, and Ar gases enter the chamber through the main valves. In this method, after entering different gases into the chamber to deposit DLC nanostructures, a plasma space is created and chemical reactions take place in a plasma space. A graphite source with a diameter of 2 inches and a thickness of 6 mm with a purity of 99.99% and argon gas with a purity of 99.99% was used as the sputtering gas in all layers. The distance between the samples and the cathode was set to 5 mm for all the studied samples [3, 6, 7].

After emptying the air in the chamber, the selected gas is injected into it, and by applying a voltage of 13.56 MHz, plasma is formed between the electrodes. During this time, the intended membrane placed in the intended location is affected by the plasma. The pressure of the chamber during irradiation was fixed at 70mtorr and the power of the amorphous carbon source was set at 150W for making layers on membranes surface. In order to minimize the thermal effects caused by the plasma and damage to the membranes, the plasma pulse mode with a duty cycle of 15% was used. The total radiation time of the film is 30 minutes. After every 30 seconds of amorphous carbon irradiation on the surface of the samples, 15 seconds of argon plasma was applied (until the end of 30 minutes of surface treatment). This coating is applied with a power of 50 watts with a duty cycle of 15% to separate loose carbon bands or hydrocarbons placed on the surface so that the final created layer is a dense carbon layer and is stable in the environment.

In most cases, the deposition is linear and the coating sits only on one side of the substrate. We have 3 membrane samples, which were subjected to 3 types of irradiations for each batch: once on the membranes, the second time on the back of the membranes, and lastly on both sides of the membranes, the process is repeated twice (Table 1). In all these samples, the amount of hydrophilicity can be increased by controlling the amount of amide or carboxylic groups. For amine groups, ammonia was used, for amide groups, Oto or acetic acid was used. The best ratio for creating hydrophobic surfaces is 15ccm acetylene and 5ccm methane. Finally, after the specified time is over and the voltage is cut off. According to the latest studies, the modification of the PES membrane surface by the method of coating DLC nanostructures to simultaneously improve the anti-fouling property and the ability to separate ions has not been reported yet. In this sense, as has been said before, different states of the surface modifier layer were placed on the membrane. AFM analysis of salt separation, water flux, porosity, hole size, the CA of water, and anti-fouling properties were carried out on the modified membranes for study.

Table 1. Types of coating in the process of membrane surface modification with DLC nanostructures.

| Membrane | Coating layer |
|---|---|
| $M_0$ | Control |
| $M_1$ | One of the two sides of the surface (Down) |
| $M_2$ | One of the two sides of the surface (Up) |
| $M_3$ | Both sides of the surface (Up & Down) |

**Membrane analysis**

Infrared spectroscopy was performed with a spectrometer to identify chemical groups. To prove the formation and existence of the layer deposited on the surface of the PES nanofiltration membrane, the analysis of the scanning electron microscope images was used. To evaluate the change in hydrophilicity, the water CA technique was used. In this sense, water without ions was dripped on the above membranes at three points randomly with drops and the average results were reported. To study the roughness of the surface and its change after the coating process, atomic microscope (Femto-Scan) images of the surface were prepared. The percentage of porosity and the amount of the size of the holes were used from the following relationship to calculate the percentage of porosity.

$$\varepsilon(\%) = \left(\frac{W_w - W_d}{\rho_f V_m}\right) * 100 \tag{1}$$

Here, $W_w$, $W_d$, $\rho_f$, and $V_m$ respectively represent the wet and dry weight of the membrane (gr), water density (gr/cc), and the volume of small pieces of the membrane (cm³). To ensure the reduction of error, the measurements were repeated three times and the average values were reported. The size of membrane holes was also calculated from the following relationship [38]:

$$r_m = \sqrt{\frac{(2.9 - 1.75\varepsilon)8\eta/Q}{\varepsilon A \Delta P}} \tag{2}$$

Where, $\eta$, $Q$, and $\Delta P$ respectively are the viscosity of water ($8.9 \times 10^{-4} Pa.s$), the volume of pure water flux passing through the membrane (m³/s), and the operating pressure ($0.55 MPa$).

**Flux and membrane separation performance**

Examining the performance of modified membranes is similar to previous researches [7]. Pure water was used to compress the membrane at a pressure of 0.6Mpa for 20 minutes. Then the operating pressure was set to 0.55Mpa for the salt test.

$$j_{w,1} = V/{A\Delta t} \qquad (3)$$

Pure water flux was calculated from the following equation, in which the flux is L/m²h, V is a liter, A is square meter, $\Delta t$ is a hour. Sodium sulfate salt solution with a concentration of 0.01 m/L was used as a feed solution to check the performance of the prepared cells. The following relationship was used to measure the separation rate.

$$R\% = \left(\frac{C_f - C_p}{C_f}\right) * 100 \qquad (4)$$

In the above relationship, $C_p$ and Cf are respectively the amount of salt concentration in the water permeated from the membrane and the feed.

**Fouling test**

In order to investigate the clogging properties of the studied membranes, the closed-end tube was immediately filled with (8000 mgr/L) milk solution. Based on the amount of water seeped from inside the membrane, the protein solution $J_P$ (L/m²h) was measured at a pressure of 0.55 MPa for 120 min. Deionized water was used to wash the captured membrane after blocking it with milk. To estimate the water recovery rate, the net water flux was measured again after the milk test process and was named $(J_{w,2})$ L/m²h. The flux recovery rate ($FRR$) was calculated from the following relationship.

$$FRR\% = \left(J_{w,2}/J_{w,1}\right) * 100 \qquad (5)$$

Likewise, for measurement of the rate of total membrane eclipse the following relationship was used [33, 39]:

$$Rt(\%) = \left(1 - J_n/J_{w,1}\right) * 100 \qquad (6)$$

**Discussion and results**

**Infrared spectroscopy and membrane hydrophilicity**

Figure 1 shows the Fourier transform infrared spectroscopy image of the bePES membrane and PES membrane coated with DLCs. In the spectrum related to the PES membrane, the stretching peak of the benzene ring is observed at 3091 cm$^{-1}$ [40]. C-H bending absorption bonds of the benzene ring are seen at 718, 797, and 836 cm$^{-1}$. The three peaks that occurred at 1405, 1485, and 1578 cm$^{-1}$ are related to the aromatic skeletal vibration of the C=C double bond [41]. Absorption bonds at 1239 and 1322 cm$^{-1}$ are created due to the existence of a C-O-C stretching bond. Also, the stretching peak of the S=O bond at 1149 cm$^{-1}$ can be seen in the structure of the PES, and the bond at 1738 cm$^{-1}$ is a sign of the symmetrical stretching of the C-O bond related to the carboxyl anion. The absorption bond located at 3115 cm$^{-1}$ also the C-H stretch of aromatics shows [40, 41]. According to the figure, the Fourier transform infrared spectrum of the modified PES membrane with DLC nanostructures and the pure PES membrane are very similar, and the peaks of the two samples almost coincide. However, at 1738cm$^{-1}$, we see an increase in the intensity of the absorption peak, which indicates the C-O stretching bond, which confirms the presence of DLCs. This means that if it is possible, a part of the DLC layer has been washed and removed from the membrane, but a part of it still remains on the surface of the membrane. It seems that the effect of DLC coating on the physical characteristics of the PES membrane surface is more than its chemical structures.

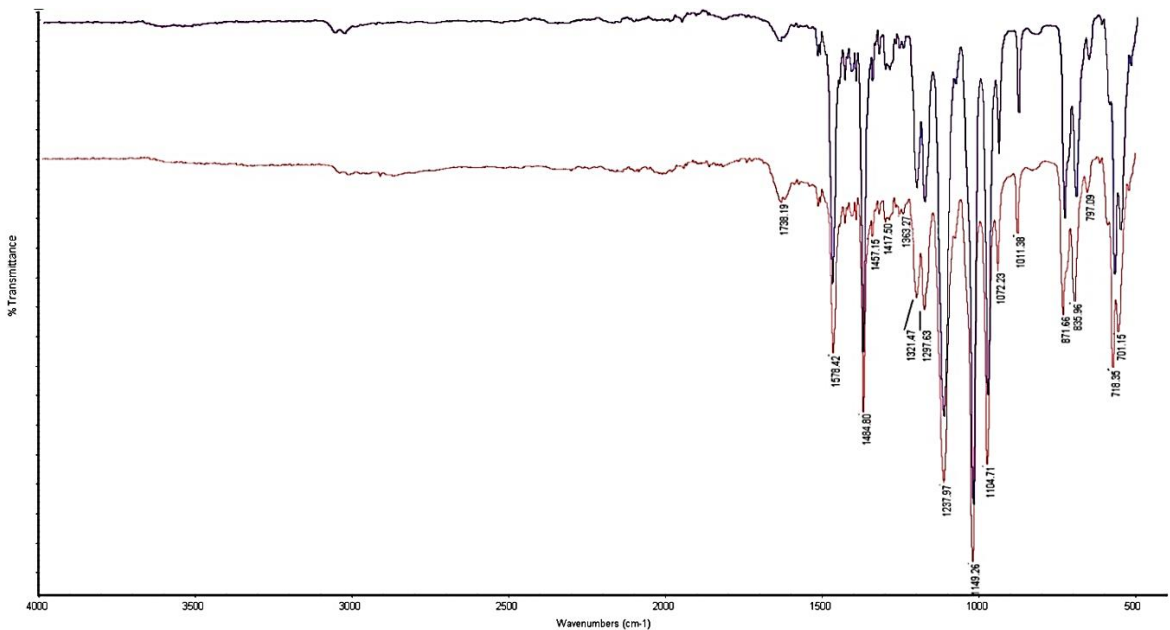

Fig 1. Fourier transform infrared spectroscopy image of the control PES membrane (up) and membrane modified with DLC nanostructures (down).

The level of hydrophilicity of the membrane surface is a parameter that can be estimated by measuring the CA of water. Figure 2 shows the measured water CA for the membranes before and after the modification process. To check the wettability of the studied surfaces, it is enough to analyze the results of different analyses. In general, by reducing the water CA parameter with the membrane surface, the wetting property or in other words, the hydrophilicity of the modified surfaces increases in proportion to the surface of the control membrane [42]. Therefore, the CA of the water drop on the surface of the PES control membrane has the highest value (84°C). Following the coating of DLC nanostructures on the surfaces of hydrophobic PES membranes, a decrease in the contact angle of water drops on these surfaces was observed (42°C). Since the contact angle of the modified samples is lower than the control sample, it can be concluded that the hydrophilicity increases. In this case, it can be said that the improvement of the hydrophilicity of the modified membranes is due to the carboxyl active groups.

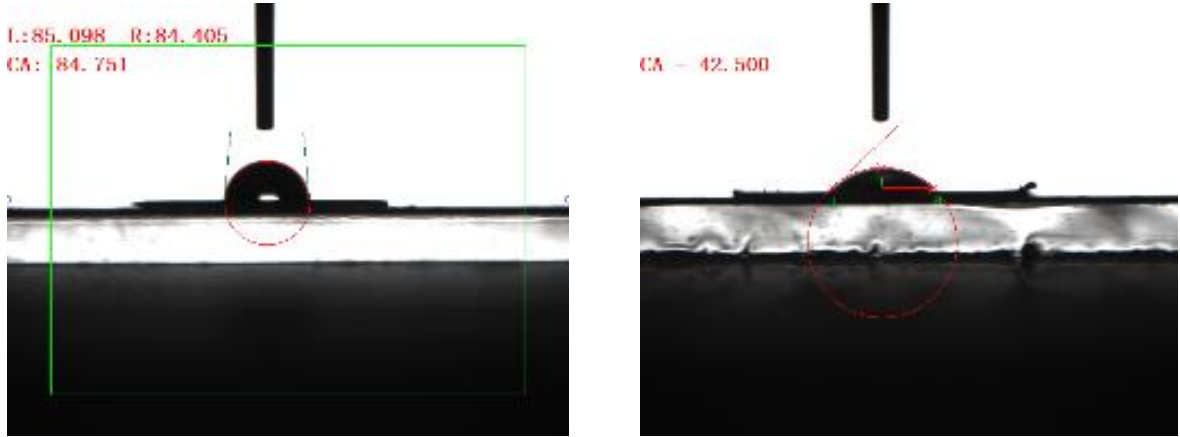

Fig 2. The CA images of the control PES membrane (Left) and membrane modified with DLC nanostructures (Right).

**Electron and atomic microscope images**

Scanning electron microscope images of raw and modified membrane surfaces were prepared and shown in Figure 3. This analysis was done to check the presence of a DLC nanostructure layer on the surface of the membrane. Based on this image, the formation and placement of a thin layer on the surface of the PES membrane is clearly evident, which is an indicator of the success of the coating process.

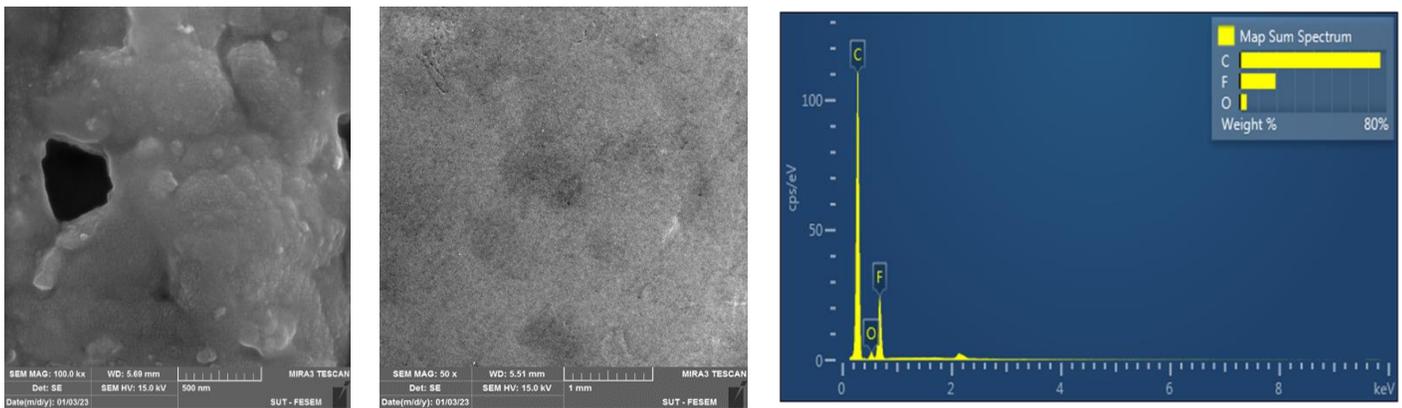

Fig 3. The SEM images of the control PES membrane (Left) and membrane modified with DLC nanostructures (Right).

In order to investigate the physical structure of the surfaces of the modified membranes as well as the control membrane, atomic microscope images were studied. The morphological changes of the surface of these membranes and their roughness due to the etching process were shown with atomic microscope images in Figure 4 and Table 2. In the case of membranes modified with DLC nanostructures, the deposited layer uniformly covers the ridges and ridges on the surface and creates a smoother surface compared to the unmodified PES membrane.

| membrane | $R_a$ ($\mu$m) |
|---|---|
| $M_0$ | 16 |
| $M_1$ | 9 |

Table 2. Average Roughness of all membranes.

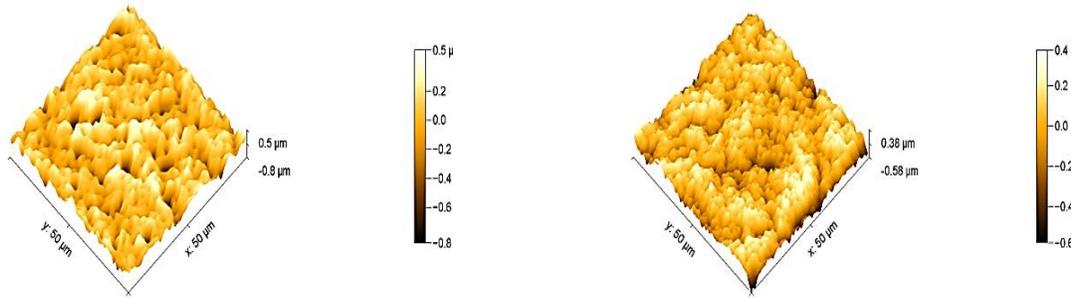

Fig 4. The AFM images of the control PES membrane (Left) and membrane modified with DLC nanostructures (Right).

## Separation function

Figure 5 shows the pure water flux through the studied membranes. As can be seen, the amount of water flux in the unmodified PES membrane is several times lower than in the coated membranes. As seen in the atomic microscope images, the deposition of DLC nanostructures on the surface of the PES membrane led to the production of membranes with a smoother surface. It is obvious that the existence of this thin layer on the surface of the PES nanofiltration membrane will cause surface holes to be covered and blocked, and as a result, the size of these holes will be reduced (the results of hole size and porosity in Figure 6). This thin layer on the surface of the membrane plays a role as a parameter in increasing the resistance to mass transfer (against the passage of water through the membrane). Since the results obtained from the water CA indicate an increase in hydrophilicity on the surface of the modified membranes, it is expected that this increase in hydrophilicity will lead to an improvement in the water flux.

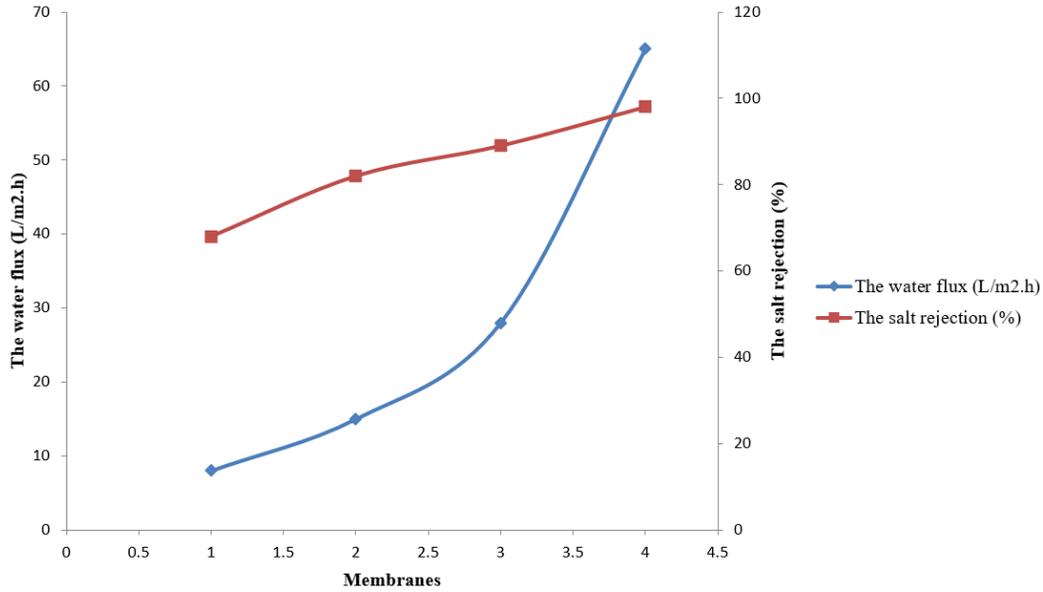
Fig 5. Effect of coating of DLC nanostructures on water flux and salt separation percentage.

In addition to the fact that the amount of water flux increased significantly with the coating process on the surface of the coated membranes, the positive effect of this process is also noticeable in other parameters, and it can be said that compared to the process of producing a network membrane mixed with nanoparticles, the amount of nanostructure used in this method is much less. The results obtained for the percentage of salt separation by the membranes before and after the layer process are reported (Fig 5). According to the results, it can be seen that the separation ability of PES modified membranes was 98% (M3) and 64% in the control membrane (M0), so it can be said that this ability has clearly increased due to plasma modification. The increase in rejection occurred due to the surface absorption of DLC nanostructures by the surface cavities of the PES membrane [43].

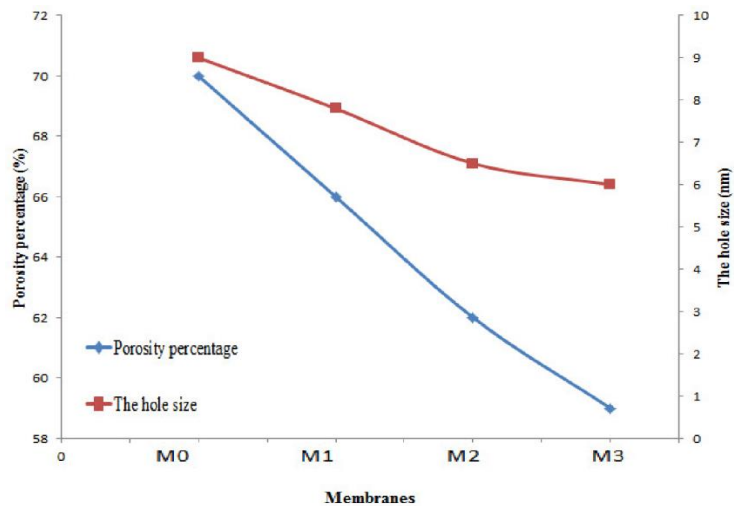
Fig 6. The hole size and porosity percentage for the studied membranes.

The presence of a negative charge on the surface of DLCs due to the active carboxyl group can enhance the electrostatic attraction of positively charged sodium ions. In this way, there is an increase in repulsive forces between negative charges (sulfate ions) and the membrane surface, which is a direct effect of the presence of active carboxyl groups [44]. In addition, in the case of membranes modified by DLC nanostructures, due to the high absorption property of DLCs nanostructures, the amount of absorption power of transient ions from the membrane increases [45]. Therefore, it can be said that after the surface modification process, the amount of the PES radiation absorption property and as a result the amount of separation percentage increases visually. Coating on both sides of the membrane surface increases the number of active adsorption sites on the membrane surface. As mentioned in the section related to net water flux, the amount of water flux increased by increasing the coating on both sides of the membrane surface with DLCs. Increasing the amount of water flux means less contact time between the feed solution and the membrane surface for water to pass through the membrane and can help to speed up the ion adsorption process by the adsorption and separation sites. Based on the observed results for pure water flux and separation percentage, the modified membrane only from the bottom side shows the least amount of improvement in separation percentage and the least amount of increase in water flux. If the goal of coating is to obtain a membrane with high water flux and ion separation ability, it seems that the modified membrane on both sides with good performance is the best membrane to choose. This membrane is the best membrane among the flat reformed membranes (highest flux and significant salt separation percentage) with suitable performance in flux and separation.

**Anti-fouling properties**

Among the important factors that severely limit the use of membranes in separation processes is their fouling, because the efficiency and lifespan of the used membrane decreases. The deposition or absorption of materials in water, such as solid particles, mineral salts, etc., on the surface of the membrane or inside the holes is called fouling. Among the adverse effects of the sedimentation phenomenon, we can mention the reduction of the flux passing through the membrane and its low performance in the separation process [1]. The amount of flux through the membrane for pure water before and after the separation of milk solution and also the flux of water and milk solution over time for PES membrane and membrane M3 is shown in Figure 7.

The parameters of flux recovery rate and total absorption rate for uncorrected PES membrane and membrane M3 as the optimal leveled membrane were investigated and evaluated and the result of this investigation is in Figure 8. As can be seen, for both membranes, a decrease in flux has occurred due to the protein particles settled on the surface of the membrane. In the first step the amount of water flux for the PES membrane is more than that for the membrane M3, which is related to the existence of a correction layer on the modified membrane and greater resistance to water penetration (it was fully explained in the previous section).

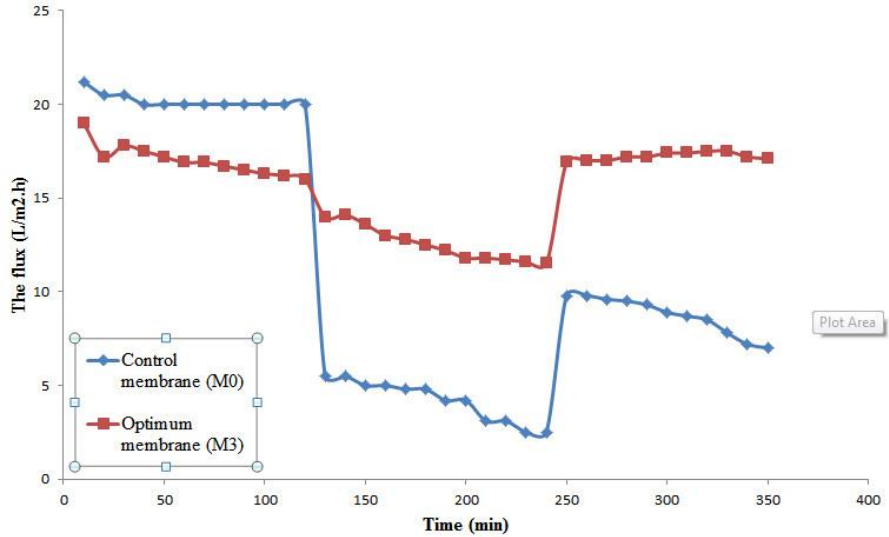

Fig 7. The amount of pure water flux before and after the milk solution separation process over time.

The amount of protein flux (Jp) was several times higher for the modified membrane (membrane M3) compared to the non-modified membrane (PES) in this graph. The third part of this diagram in Figure 7 surprisingly showed a lower amount of flux after the process of washing the protein-clogged membrane with water for the PES membrane compared to the modified membrane. In Figure 8, it can be clearly seen that the value of the water flux recovery rate for membrane M3 is about 91%, which is more than doubled compared to its value for the PES (40%) membrane. These results indicate a direct relationship between the flux recovery rate and the level of hydrophilicity of the prepared membranes. Improving the rate of hydrophilicity of the membrane and, as a result, preventing the deposition and loading of substances and factors that cause fouling on the surface of the membrane is a very acceptable and well-known thing. In this case, the deposition and fouling of materials on the surface of the membrane occur with less intensity, and the removal of these materials from the surface of the membrane (as a fouling agent and protein) is easier with the process of washing with water after filtering the milk solution [45].

Therefore, the improvement of the anti-fouling properties of membrane M3 can be attributed to the improvement of the level of hydrophilicity of the surface of this membrane. In addition, the smoother surface of this membrane in comparison with the unmodified sample (PES) is also known as another factor in reducing the seizure phenomenon [30]. The obtained results for the parameter of the total capture rate reported are in complete coordination and compliance with the results related to the flux recovery rate in Figure 8. In any case, the amount of total absorption in membrane M3 significantly decreased from 76% to 30% in comparison to the uncorrected membrane. This reduction means the greater resistance of the modified membrane with DLC nanostructures against the blocking factors.

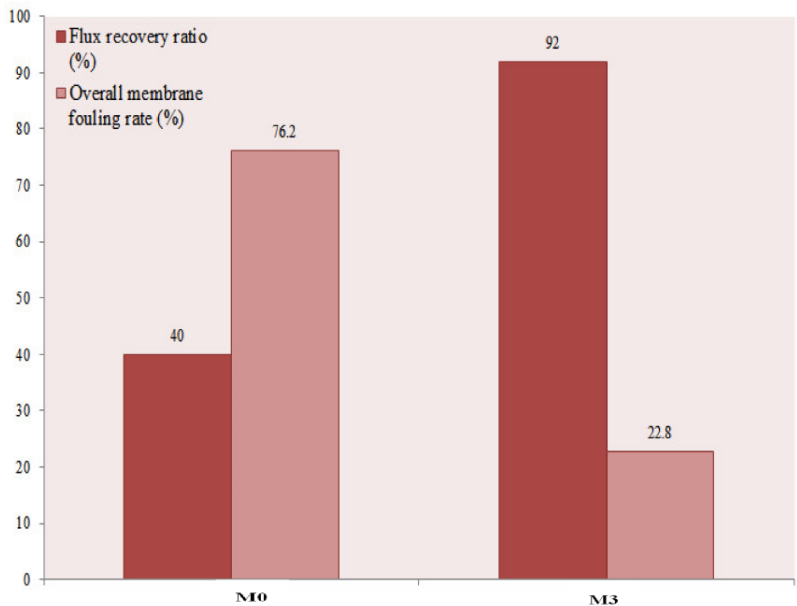

Fig 8. Comparison of the water flux recovery rate and overall clogging rate for the PES membrane and M3 sample.

**Conclusion**

In this research, the improvement of the surface of a PES hydrophobic nanofiltration membrane was considered using a new method, which is the coating of DLC nanostructures. For this purpose, the coating of each of the upper and lower surfaces (separately and simultaneously) of this membrane was studied. The water CA, atomic microscope images, and parameters of surface roughness, water flux, salt separation percentage, hole size and porosity percentage, and anti-fouling properties of the membrane were studied and discussed. Due to the placement of a thin correction layer on the surface of the polyester sulfone membrane, the amount of trans-missive current increased, and the size of surface holes and the percentage of porosity also decreased. This was while the percentage of salt separation was greatly improved from 64% to 98% with the index layer. The amount of water CA as a measure of membrane hydrophilicity showed a noticeable improvement in the effect of the index layer. Evaluation of the surface of the membranes with atomic microscope images showed a reduction in the surface roughness of the modified membranes. The results showed that the value of the flux recovery rate for membrane M3 (membrane on both sides of the layer) as the optimal membrane was many times higher than that for the PES membrane. Similarly, the amount of the total absorption rate of membrane M3 was lower than that of the pure membrane. According to the obtained results, it can be claimed that the coating of DLC nanostructures on the surface of the base membrane is a new and suitable method for the development of these membranes with improved anti-fouling properties.

It should be noted that this test was also performed on PP and PVDF membranes in addition to the PES membrane, but since the PES membrane had the best results, we reported it. In the end, the modified membranes by the above method, were immersed in a saturated salt solution for 48 hours and then dried. It was observed that the least amount of sediment remained in the PES membrane modified by DLC nanostructures and the salts attached to the surface were easily removed from it. This was the surprising goal we were looking for (videos are attached to compare the results from all 3 membranes). The coated sample has a hydrophilicity property after washing up to 10 times or more. These results indicate that the created film has maintained its stability by increasing the number of washing cycles. After each washing, the samples are dried for 15 minutes at 60 degrees Celsius. This drying period is another opportunity for PES to reorganize the broken bonds, and it is because the samples have a low contact angle after drying. Silicone polymers have the property that they can reorganize themselves after breaking bonds. Based on the washing test, the prepared sample has good stability against the number of washing cycles, which is a very important advantage for practical applications.